\documentclass[12pt,preprint]{aastex}

\shorttitle{SN2006jc}  
\shortauthors{Di Carlo et al.}

\begin{document}  
  
\title{Near-Infrared observations of the type Ib Supernova SN2006jc:
        evidence of interactions with dust}

\author{E. Di Carlo\altaffilmark{1}; C. Corsi\altaffilmark{2};   
A. A. Arkharov\altaffilmark{3}; F. Massi\altaffilmark{4}; 
V. M. Larionov\altaffilmark{3,5}; N. V. Efimova\altaffilmark{3}; \\
M. Dolci\altaffilmark{1}; N. Napoleone\altaffilmark{2}; 
A. Di Paola\altaffilmark{2}}

\email{dicarlo@oa-teramo.inaf.it}

\altaffiltext{1}{Istituto Nazionale di Astrofisica - Osservatorio Astronomico 
Collurania-Teramo, Via Mentore Maggini I-64100, Teramo, Italy}  
\altaffiltext{2}{Istituto Nazionale di Astrofisica - Osservatorio Astronomico 
di Roma, Via Frascati 33, I-00040, Monteporzio Catone (Roma), Italy}  
\altaffiltext{3}{Central Astronomical Observatory at Pulkovo, Pulkovskoe 
shosse 65, 196140, Saint Petersburg, Russia}  
\altaffiltext{4}{Istituto Nazionale di Astrofisica - Osservatorio Astrofisico 
di Arcetri, Largo Enrico Fermi 5, I-50125, Firenze, Italy}  
\altaffiltext{5}{Astronomical Institute of St. Petersburg University, Russia}
  
\begin{abstract}  
In the framework of a program for the monitoring of Supernovae in the 
Near-Infrared (NIR) carried out by the Teramo, Rome and Pulkovo observatories 
with the AZT-24 telescope, we observed the Supernova SN2006jc in the  $J,H,K$ 
photometric bands during a period of 7 months, starting $\sim  36$ days after
its discovery. Our observations evidence a NIR re-brightening, peaking $\sim 
70$ days after discovery, along with a 
reddening of $H-K$ and $J-H$ colors until 120 days from discovery. After that 
date, $J-H$ seems to evolve towards bluer colors. Our data, complemented by 
IR, optical, UV and X-ray observations found in the literature, show that the
re-brightening is produced by hot dust surrounding the supernova, 
formed in the interaction of the ejecta with dense circumstellar matter.
\end{abstract}

\keywords{supernovae: general --- supernovae: individual (SN~2006jc) --- 
galaxies: individual (UGC~4904) --- infrared: stars}

\section{Introduction}  
One of the main aims in the astronomical research on Supernovae (SNe) has been 
the
study of hydrogen-deficient objects, among them Type Ib SNe (helium-rich) and
Type Ic SNe (helium-deficient; see Filippenko 1997 for a review on SNe classification). 
These two subclasses of events are of great interest because, more than for
all other 
types into which SNe have been classified, there are still
many issues open about 
them, both in terms of observational properties and concerning their 
progenitors and the physics of the explosion. A number of different models 
have been proposed to account for
the observed characteristics of SNe Ib/c. The most agreed scenario is that of
the core-collapse (CC) of massive stars. It is now believed that their progenitors
are Wolf-Rayet stars, i. e., highly evolved stars that have shed their H 
envelopes
(or He envelopes in the case of SNe Ic progenitors) due to mass transfer in a
binary system or by means of a stellar wind (for an updated review,
see Crowther 2007). However, no observational
evidence has been found to date allowing one to identify an explosion
picture.   

Since supernovae are believed to be a major source of dust in galaxies 
(Todini \& Ferrara 2001, and references therein),
events showing signatures of interactions with circumstellar matter
are of great interest to study how dust forms around these objects.
In particular, NIR observations until late epochs are instrumental
in providing information on the ongoing processes. As an example,
IR traces of dust condensation were found in the ejecta of SN1987A
starting $\sim 500$ days after the outburst (see, e.\ g., Bouchet \&
Danziger 1993). Unfortunately,
only in few cases good NIR light curves spanning until late epochs are 
available.  
The observatories of Teramo (Italy), Rome (Italy) and Pulkovo (Russia) are 
involved in a
program aimed at providing accurate light curves of SNe in the Near-Infrared
(Di Carlo et al.\ 2002, Di Paola et al.\ 2002, Valentini et al.\ 2003) by 
using a dedicated 1m class telescope equipped with a Near-Infrared camera.
In the framework of this program, 
we present the results of our NIR follow-up of SN2006jc, 
discovered on 2006 October 9.752 UT (Nakano et al. 2006) in the Galaxy UGC 
4904. 
The earliest photometric and spectroscopic observations (Crotts et al. 2006; 
Itagaki et al. 2006; Modjaz et al. 2006) already identified the object 
as a peculiar Type Ib SN a few days past maximum light.
The main feature of SN2006jc in the NIR is a late re-brightening in the 
corresponding Light Curves (LCs), making SN2006jc an intriguing event, that 
gives us a huge opportunity for a study of dust formation around  CC SNe and 
allows to deepen our knowledge on how massive stars die.

Nakano et al.\ (2006) soon drew attention to an LBV-like luminous flare that 
had been detected in the same position as SN2006jc two years before. The 
spatial coincidence of the two events was carefully analyzed by Pastorello et 
al.\ (2007). 

Both the character of the event and the preceding flare occurrence,
led many groups to carry out ground-based and satellite observations from
many facilities and over a wide range of wavelengths, providing a wealth
of data.
Arkharov et al.\ (2006) were the first ones to note that the NIR brightness of 
the SN was undergoing a steady increase after the optical maximum, suggesting 
some kind of relation with dust.
Immler, Modjaz \& Brown (2006) also suggested this, based on X-ray
observations.
It can be hypothesized that as a result of the LBV-like flare in 2004, a dense circumstellar medium was originated within which the supernova 
exploded. However, LBV stars are supposed to be massive stars in a brief and 
unstable evolutionary phase from core H-burning to He-burning (WR stage). So, 
stellar evolution theories do not predict that a star in this phase can become 
a Supernova after only two years or that a WR star would produce such a 
luminous eruption (Eldridge \& Tout 2004; Heger et al. 2003). 
Pastorello et al.\ (2007) suggest as an alternative picture that 
the SN progenitor was part of a binary system and that the LBV-like eruption 
actually occurred in the companion star of the SN2006jc progenitor. 
Nevertheless, the SN would have exploded  in a dense CSM, as testified by the 
evidence of interactions with it.
Tominaga et al.\ (2007) used the available observations from the X-ray to the
optical, complemented with our NIR photometry,
and modeled the event as the explosion of a WR star that
was $\sim 40$ $M_{\odot}$ during its main sequence phase. Dust 
carbonaceous grains would then have formed within the ejecta.

This is one of the peculiarities of SN2006jc, since dust signature is hardly 
found in Type Ib/c SNe. One of the few instances of Ib SNe whom dust may
have been associated with is SN1990I. However, in this case the presence of 
dust was only indirectly inferred from the optical data (missing NIR
information) around day 250 (Elmhamdi et al.\ 2004).
SN2006jc is probably the very first instance of a Type Ib SN showing clear 
evidence of interaction with a dusty circumstellar 
environment, be the dust newly-formed or pre-existing.

The paper is organized as follows. 
We first describe the observations and data reduction (Sect.~\ref{obse}). 
Then we present the NIR light curves and color evolution of SN 2006jc, 
and describe the construction of a bolometric (uvoir) LC (Sect.~\ref{resu}). 
Finally, in Sect.~\ref{disc:conc} we discuss the most likely origin of the NIR 
emission and its main constraints to the circumstellar medium.

\section{Observations and data reduction}  
\label{obse}

The observations of the SN 2006jc were performed with the 1.1m AZT-24 
Telescope at the Campo Imperatore  
Observatory (Italy), equipped with the Near-Infrared SWIRCAM camera (Brocato 
\& Dolci 2003), which is based on a $256 \times 256$ HgCdTe NICMOS3-class array (PICNIC).   
The detector is sensitive  
to radiation in the spectral range from 0.9 to 2.5 $\mu$m and, at the focus of AZT-24,  
yields a scale of $1.04$ arcsec pixel$^{-1}$, resulting in a field of view   
of $4.4 \times 4.4$ arcmin$^2$.  
  
A single set of observations in one of the $J$,$ H$,$ K$ bands 
consists of 3--5 off-source images and 5 
on-source images. During a night, for each band the number of sets acquired 
was selected based on the source brightness in that band.   
The on-source images in a same set are dithered with maximum shifts of $15"$  
and the individual integration times are 30 seconds.  
Off-source frames are offset by $20'$ and were taken with the same observational procedure  
and exposure times as the on-source ones, but using a wider ($30"$) dithering.   
For each set, sky images were obtained as a median of the closest  
dithered off-source frames and subtracted from each on-source image in the 
same set.   
  
A number of frames were acquired at twilight and used to obtain
normalized flat-field images using the differential flat  
technique, as described in Di Carlo et al.\ (2002). Each sky-subtracted
image was flat-field divided accordingly. 
To correct for residual low spatial frequency inhomogeneities in the system response over the frame,  
we used a set of observations of a standard star with many different pointings   
covering a grid of $7 \times 7$ regularly-spaced positions on the detector area  
plus 7 more positions at the center.  
For each of the 56 frames, a "sky" image was constructed by median-filtering  
together the 12 frames closest to it in time. Each sky was then removed from 
the corresponding frame and the obtained images were corrected for flat-field. 
Aperture photometry of the standard star was carried out on each image,  
yielding a flux (in counts) as a function of the position on the grid. We  
fitted a polynomial of 2nd degree to the grid and transformed it into a frame  
that, once normalized to the value at the centermost location, was used as a   
"super-flat". Therefore, after flat-fielding, each frame was further divided 
by the super-flat. The super-flats evidence large-scale smooth variations of at most $\sim 0.1$ mag from the frame center to its edge.   

At last, the relevant scientific images were then obtained by removing the bad pixels, 
and finally registering and averaging together the dithered exposures using a 
$5 \sigma$ threshold. 
  
\begin{figure}  
\includegraphics[angle=-90,scale=.70]{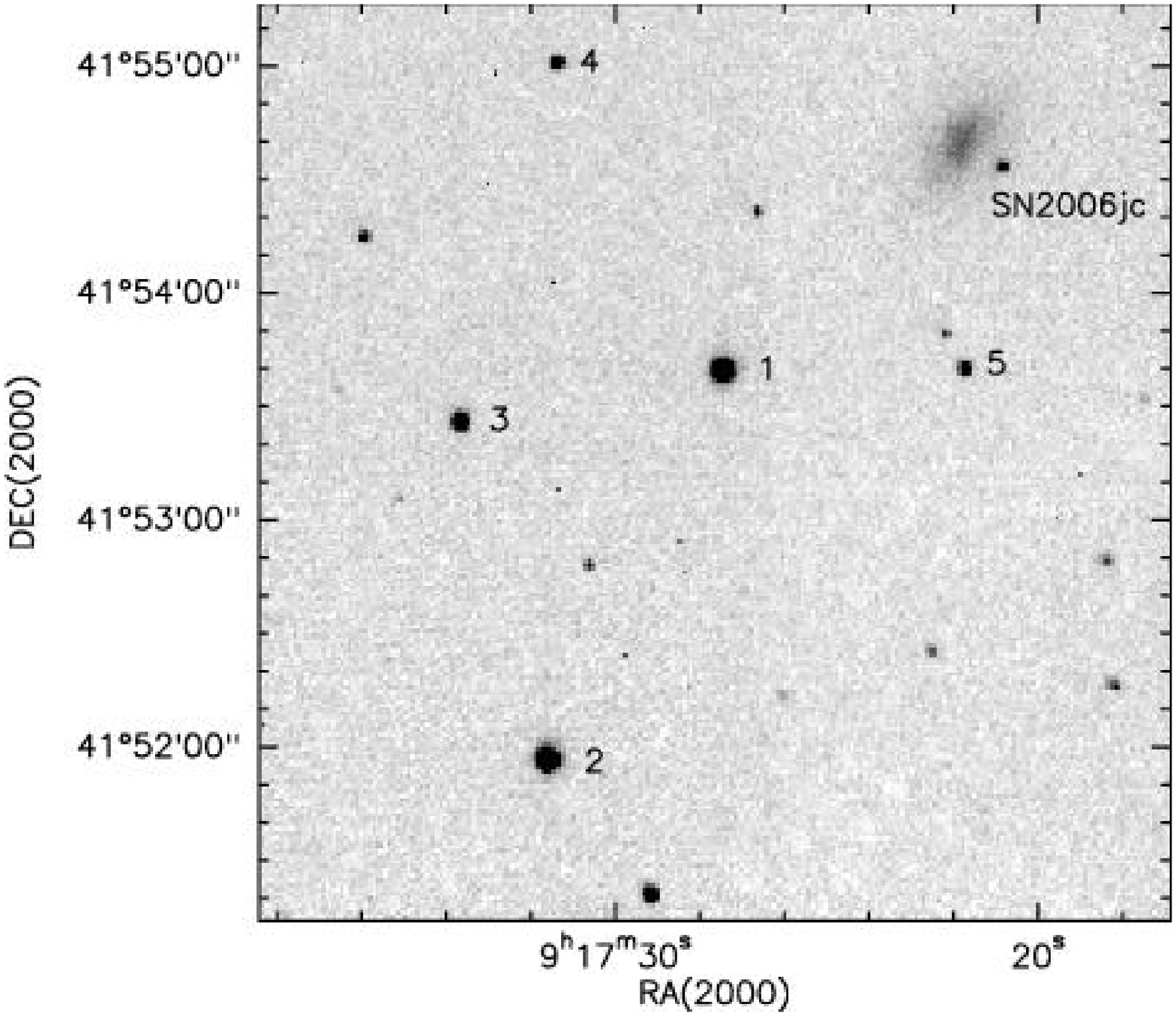}  
\caption{J-band image of SN2006jc obtained at the AZT-24 telescope on JD 2,454,083. Stars of the   
local sequence are indicated by numbers. \label{fig1}}  
\end{figure}

Photometric observations in the $JHK$ bands began on November 15, 2006, 
going on until June 6, 2007.  
The standard star AS20 (Hunt et al.\ 1998) was observed to calibrate a sequence 
of secondary standards properly selected in the supernova field among the ones 
with the highest S/N (stars 1--5 in Fig.~\ref{fig1}). 
We checked the calibrated magnitudes of the stars in the local sequence 
against the 2MASS catalog and found that they are in general
within $\sim 0.02-0.04$ mag of their 2MASS magnitudes.
Only the two faintest ones exhibit larger differences in some of the band.
Also, we found that the relative magnitudes of the local sequence remain 
constant within a few $0.01$ mag.

\begin{figure}  
\includegraphics[angle=0,scale=.80]{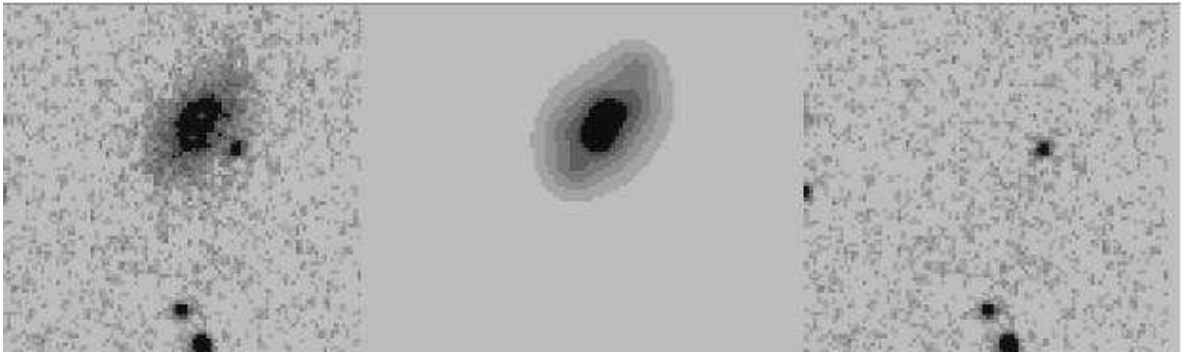}  
\caption{Sketch of the procedure used to remove the background contribution
from the SN photometry.  
The region around the SN in the original image is shown in the left  
panel and the same region after background subtraction in the right panel.
The subtracted contribution from the host galaxy
has been modeled as a sum of 6 gaussians (middle panel).  
\label{fig2}}  
\end{figure}

The magnitudes of SN2006jc and the 
secondary standards in the field, were obtained by Point-Spread Function 
photometry, using the ROMAFOT package (Buonanno et al. 1979). The procedure 
adopted to remove the 
contribution of the galaxy background diffuse emission is sketched in 
Fig.~\ref{fig2} and consisted of modeling the galaxy by a set of several 
(5--8) gaussians with variable Full-Width at Half-Maximum.
The analytical reconstruction of the galaxy was subtracted from the original 
image and the background residual around the SN was checked interactively. 
Finally, the SN photometry was carried out on the subtracted image showing the 
flattest background distribution.
Table~\ref{table1} lists the results of our photometry. The uncertainties due
to photon statistics and fitting are also indicated.

\section{Results}  
\label{resu}

\subsection{The Near-Infrared Light Curves}  
\begin{figure}  
\includegraphics[angle=-90,scale=.70]{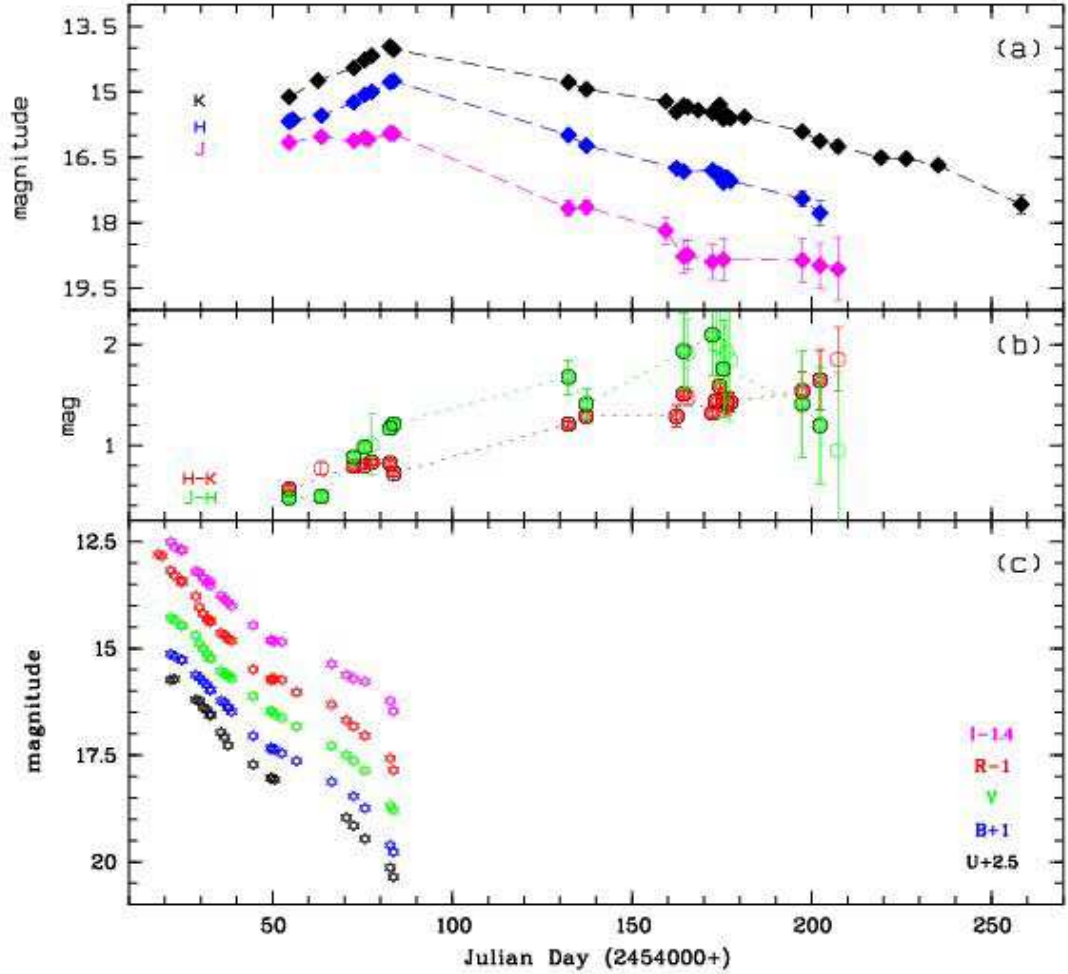}  
\caption{(a) NIR light curves and (b) the corresponding NIR colors ($J-H$ and 
$H-K$) of SN2006jc. Full circles correspond to epochs when photometry in all 
three bands is available, whereas empty circles mark epochs when only 
photometry in two bands is available and this has been complemented by 
interpolated values in the missing band. (c) To allow a more complete view of 
the NIR temporal evolution, the optical light curves (from Pastorello et al.\ 
2007) are also shown, shifted by the amount indicated on the lower-right 
corner of the box.\label{figLCs}}  
\end{figure}

The $JHK$ Light Curves of SN 2006jc are shown in Fig.~\ref{figLCs}. 
Unfortunately, no 
NIR data are available around the optical maximum, the coverage starting 
roughly 36 days after discovery. 
The NIR LCs span a period of about 7 months and exhibit a clear peak (hereby, 
the re-brightening) around JD2454083 or shortly after (i. e., 60--70 days 
after discovery) 
that is correlated with a slope decrease in the light curves at optical
wavelengths (see Fig.~\ref{figLCs}). In $\sim 30$ days, 
$H$ and $K$ became brighter by at least 1 mag, whereas in $J$ only a plateau 
is evident (or a small brightness increase, $\sim 0.2$ mag).  Hence, the 
re-brightening 
mostly affected the NIR wavelengths and its effects became smaller and smaller
at shorter wavelengths. Then, while $J$ and $H$ faded by $\sim 3$ mag in
120 days, $K$ dimmed by $\sim 3.6$ mag in 176 days. Note that the SN
was brighter in $K$ than it was in $J$ and $H$, so could be followed longer in this band. Two optical spectra taken around the re-brightening
(Smith et al.\ 2007) exhibit a red continuum (along with a blue emission) 
that confirms excess emission in the NIR. 
  
\subsection{NIR color evolution} 
\label{ccd:text}
The evolution of the NIR colors ($J-H$ and $H-K$) is also shown in 
Fig.~\ref{figLCs}.
A reddening with time is evident, with $H-K$ having steadily increased by $\sim 1$ 
mag in $\sim 150$ days, whereas $J-H$ increased by $\sim 1.6$ mag in $\sim 120$
days and then decreased by $\sim 1$ mag, although with larger errorbars, in the 
last $\sim 30$ days of the plotted color curve.
The evolution of NIR colors can be better studied in a color-color diagram,
where it can be compared with cooling and extinction of known stellar
sources.
As shown in Fig.~\ref{fig3}, the NIR colors of the SN mainly evolve along a 
track that is in practice parallel to the reddening vector. However, at the 
latest epochs plotted in the diagram, the blueing of $J-H$
evidenced above causes the SN colors to move in a different direction,
roughly perpendicular to an extinction vector.
Although the errors in $J-H$ become large at these late
epochs, this change in color evolution might be real.
Anyway, we are mainly concerned with the color evolution in the earlier phase 
shown in our diagram, i. e., during the re-brightening time, when the photometric errors are smaller.

This behavior in the NIR colors during the re-brightening is not only totally at odds with that of other non-stellar objects like SNe Ia (e. g., Valentini et 
al.\ 2003), where dust is expected to play quite a marginal role in the light 
emission, but also different, e. g., from that of a ``dusty'' core-collapse SN like 
the IIn-type 1999el (Di Carlo et al.\ 2002). In the latter case, 
dust is expected to be the main cause of the observed NIR excess and, in 
fact, the NIR colors evolve below the reddening band of the stellar
main sequence  in an $H-K$  vs.\ $J-H$ diagram. But the spread 
that it exhibits in the space of colors is much less than in the case of SN2006jc.  Indeed, this evidence points 
to the fact that most of the NIR emission of SN2006jc at the 
epochs considered in this work, is likely to arise 
due to warm dust. 

Two processes may be envisaged to explain the  
reddening trend with time of the NIR colors:  
steady formation of dust in front of a NIR source 
(which would cause the extinction to increase with time), or cooling of  
hot (i.e., $\sim 1000$K -$ 2000$K) dust 
shells surrounding the SN (irrespective whether formed in the ejecta 
or pre-existing in the circumstellar matter).    
 
As for the first scenario, i. e., steady formation of dust in front of a  
NIR source, it leads to an extinction that is
inconsistent with the optical and NIR light curves and it can be discarded 
(see Sect.~\ref{disc:conc}). Nevertheless, it allows one to roughly derive an 
upper limit to the total mass of dust around the SN in the hypothesis of a 
spherically symmetric distribution of dust forming in the SN ejecta.
From Fig.~\ref{fig3}, the track in the color-color diagram spans roughly
$A_{V} = 15$ mag and this can be roughly assumed as the maximum extinction
that dust may cause. From this, we can make an order-of-magnitude estimate of 
the dust mass involved. By using Eq.~(7--18) from Spitzer (1978), assuming 
$R_{V} = 3.1$ and a dust-to-gas mass ratio of 0.01, we obtain that the dust  
mass column density as a function of $A_{V}$ is given by: 
\begin{equation} 
\label{ccd:eq}
M_{\rm dust} \sim 2 \times 10^{21} \times A_{V} \times m_{\rm H} / 100 
\end{equation} 
where $m_{\rm H}$ is the mass of a hydrogen atom. Note that the dust-to-gas 
mass ratio is only used to derive the dust mass column density    
from the gas column density in a galactic environment; as long as the dust 
properties are the same, the above relation holds in every other environment 
irrespective of the actual dust-to-gas ratio there. 
If we assume that the dust is homogeneously distributed in a spherical volume 
of radius $R$, then the total dust mass is given by  
$(4/3)\pi R^{2}M_{\rm dust}$; we can estimate $R = v \times t \sim 10^{16}$ 
cm, where $v$ is the ejecta velocity and $t$ is the time from the 
explosion (we adopted $v \sim 10000$ km s$^{-1}$ and $t \sim 200$ days). 
The dust mass needed to produce $A_{V} = 15$ mag 
would then be $\sim 10^{-4}$ $M_{\odot}$. If the dust was concentrated in
a thin shell at distance $R$, it is easy to see that the mass would be increased
by a factor 3. Also, we can expect that the grains are smaller than
in the interstellar medium, then lowering their effective cross sections and
further raising the mass. 
 
In the framework of the second scenario, the NIR emission would be 
due to cooling dust. By assuming the dust to be homogeneously 
distributed in a plane-parallel geometry and at a constant temperature 
(although decreasing with time), from the radiative transfer equation 
we obtain
$I_{\lambda} = q(\lambda) B_{\lambda}(T) (1 -e^{-\tau_{\lambda}})$, 
where $q(\lambda)$ is the dust emission efficiency, $B_{\lambda}(T)$ is 
the brightness of a blackbody at temperature $T$ and $\tau_{\lambda}$ 
is the dust optical depth at a wavelength $\lambda$. 
We can then use the Wien approximation and see that the NIR colors are given 
by: 
\begin{equation} 
m_{\lambda_2} - m_{\lambda_1} = A(\lambda_1,\lambda_2,\tau_{\lambda_1}) 
+ 1.086 \times \frac{hc}{kT}(\frac{1}{\lambda_2} - \frac{1}{\lambda_1}) 
\end{equation} 
where every dependence on dust properties and opacity now lies in  
the term $A$ and, so, 
the dependence on $T$ is explicit. Once ``starting'' colors are 
given (for any initial $T$, they are accounted for only in
 $A$), it is obvious that the cooling ``sequence'' causes a shift roughly 
in the same direction as the reddening vector. In particular, dust 
cooling from 2000 K to 1000 K would cause $H-K$ to redden by 1.2 mag 
and $J - H$ by 1.5 mag, enough to account for the total change in NIR 
colors observed. This is also shown in Fig.~\ref{fig3}. The choice of
the temperature interval is such to roughly agree with the determinations
of grain temperatures at different epochs discussed in the following. 
Furthermore, the upper limit is determined by the dust sublimation 
temperature, that is  $\sim 2000-3000$ K (e. g., Draine \& Lei 2002).

The late color evolution, perpendicular to the reddening and to the blackbody 
sequence, cannot be explained either by extinction or by dust cooling. 
The most likely cause is a major change in the geometry of the source, in 
temperature gradients and/or in the grains properties. 
  
\subsection{Bolometric light curve}
\label{blc:text}

We used our $J, H, K$ photometry and the published $U, B, V, R, I$ photometry, 
listed in Pastorello et al.\ (2007), to construct both a ``uvoir'' (when the 
two datasets overlap) bolometric curve and a $UBVRI$ (quasi-)bolometric light 
curve. 
We assume a distance modulus of $(25.8 \pm 2.6)$ Mpc and a galactic 
extinction $E(B-V) = 0.05$, as estimated by Pastorello et al.\ (2007). 
Measurements in all bands were first dereddened according to the extinction law 
from Cardelli et al.\ (1989), adopting $R_{V} = 3.1$ that gives
$A_{V} = 0.155 $ mag.
The optical magnitudes were converted to fluxes according to Bessel (1979), 
while the NIR ones were converted according to Tokunaga \& Vacca (2005),
following the standard of the UK InfraRed Telescope. Note that ARNICA standards
were defined using UKIRT standards (Hunt et al.\ 1998).
We used spline fits to the LCs in order to have a homogeneous set of data at 
all wavelengths for each phase. The total flux at each epoch was 
computed by simply ``connecting'' the fluxes at the isophotal wavelengths and 
determining the total area. The results are listed in Table~2 and shown in 
Fig.~\ref{bolo}. 
The error on the derived bolometric luminosity is dominated by that on the
distance  and the one due to the limited band used to calculate
the total flux. As for the first, it is easy to see that it amounts to a
$\sim 20$ \% of the bolometric luminosity. Being systematic, it causes a shift 
of the bolometric light curve as a whole. The second contribution is more
difficult to estimate, but it may be quite significant and may vary with
time. However, it is likely to cause only small
deformations of the logarithmic bolometric light curve, the
major effect being a shift of the curve, as well. 

It is clear from the figure that the inclusion of the NIR fluxes 
change significantly the slope of the bolometric LC, 
their contribution increasing with time from 
a factor 1.35 to a factor 3.6 of that from $UBVRI$. 
  
Figure~\ref{bolo} also shows the bolometric luminosity obtained on 
JD2454022.6 by
adding the UV fluxes measured with Swift/UVOT (Brown et al.\ 2006);
the UV contribution amounts to $\sim 80$ \% of that from $UBVRI$. 
We roughly estimated the UV fluxes\footnote{We used zero points, wavelengths 
and count to flux ratioes as given in the Swift WEB page, at  
{\it http://heasarc.gsfc.nasa.gov/docs/heasarc/caldb/swift/docs/uvot/index.html}\\
We also assumed $ A_{\lambda} = A_{U} $ in all three UV bands to estimate the 
reddening correction for the UV flux.} 
on other dates from the figure shown
in Holland et al.\ (2007). We found that the UV to optical 
{\em plus NIR} luminosity ratio appears to remain constant with time, although 
it is also roughly equal to the ratio of UV to optical luminosity as calculated at 
the discovery time, when NIR data are not available. Anyway, 
we decided not to scale the bolometric luminosity by including a constant
contribution from UV emission, lacking a conclusive list of UV fluxes. 
The point marked in Fig.~\ref{bolo} allows one to shift the bolometric light 
curve as a whole if a constant fraction of UV contribution is assumed.

\section{Discussion and conclusions}
\label{disc:conc}

\begin{figure}  
\includegraphics[angle=-90,scale=.70]{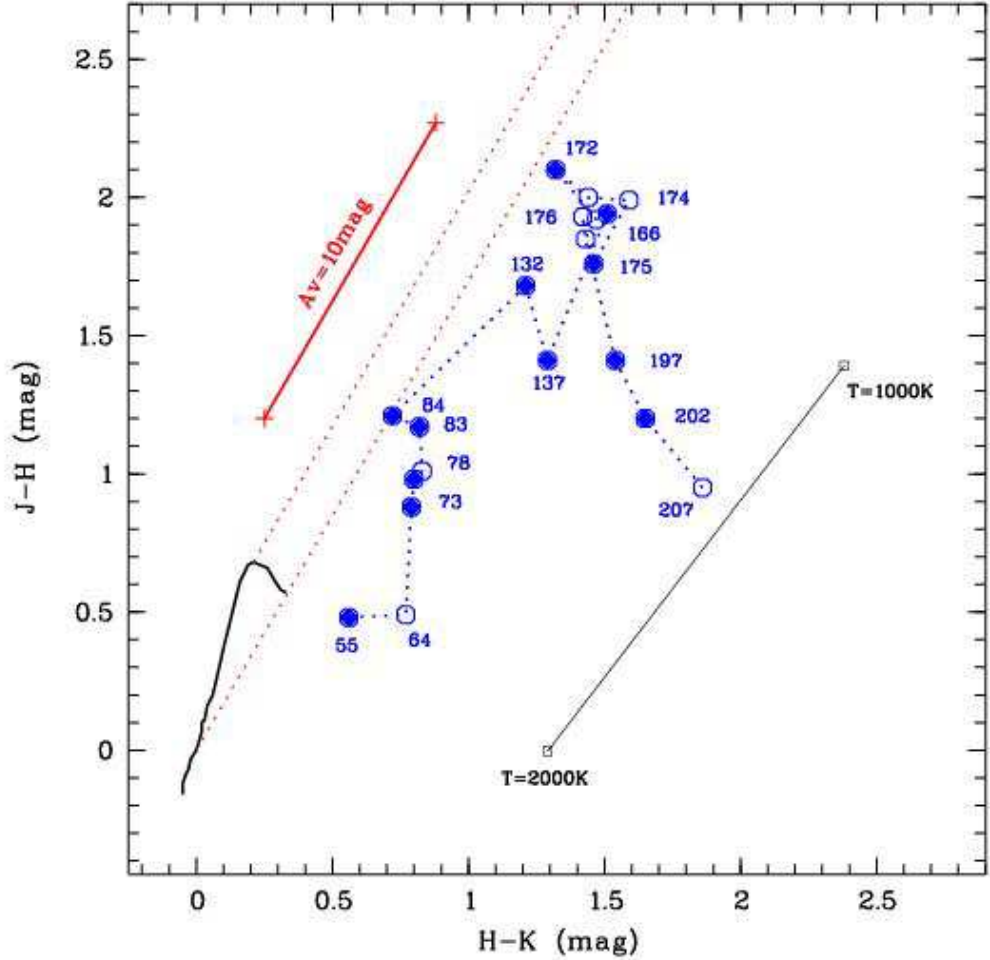}  
\caption{Color-color diagram ($J-H$ vs.\ $H-K$) showing the temporal evolution
of SN2006jc NIR colors. Each symbol is labeled with the corresponding epoch 
(JD2454000+). The symbols are the same as in Fig.~(\ref{figLCs}b). For the sake of clarity,
errorbars are omitted; however, they are plotted in the previous figure. 
The stellar main sequence of O6--O8 to M8 stars (Koornneef 1983) is also drawn 
(solid black line on the lower-left 
corner), along with its reddening band (red dashed lines).
The solid lines above and below the main sequence reddening band mark, 
respectively: an extinction $A_{V} \sim 10$ mag and a blackbody cooling 
sequence from $2000 K$ to $1000 K$. Both of them are shifted by arbitrary 
intervals.\label{fig3}}  
\end{figure}

Among the two proposed scenarios to explain the NIR color evolution, that of 
pure 
extinction can be easily ruled out. Firstly, in the time interval when optical
and NIR observations overlap, the extinction should have increased up to
$A_{V} = 5$ mag. This can be easily estimated from Fig.~\ref{fig3}, exploiting
the labeled dates. Since in this interval the increase in $V$ is ``only'' 
$\sim 2.5$ mag, 
the background emitting region should have steadily increased its brightness
to ``compensate'' for the remaining $2.5$ mag of extinction. This can be 
obviously excluded.
Secondly, even considering the much longer interval spanned by the NIR 
observations,
a pure extinction scenario can be discarded on similar grounds.
From Table~\ref{table1} it can be seen that, e. g., 
$K$ increased by $\sim 2$ mag from the epoch of the re-brightening peak to the 
latest 
dates plotted in the color-color diagram. Since on this time-scale the 
extinction $A_{V}$
should have increased by $15$ mag, that translates into 
an increase of $A_{K}$ by $ \sim 1.7 $ mag, then the growth of $K$ would be 
fully explained by the extinction. Hence, the re-brightening would have needed again a 
source in the background of the newly-formed dust whose NIR flux at first increased and 
then remained almost constant.   

The simpler way to explain the NIR re-brightening is that the 
shock or the UV-flash radiation engulfed a distant dense shell 
of circumstellar matter from a past eruptive episode of the progenitor, 
either forming new warm dust in the dense post-shock gas, or 
warming pre-existing dust. Smith et al. (2007) show that if this 
eruptive episode is related to the 2004 flare event, than the shock would  
have reached the corresponding shell in $\sim 100 - 200$ days, comparable 
with the epoch of the re-brightening.  
Whereas the bulk of the dust cooled behind, the shock or the UV flash would
have yielded new layers of hot dust. We explored this possible mechanism by solving 
the transfer equation under the same assumptions as in Sect.~\ref{ccd:text}, but  
adding a a plane-parallel outer layer of warmer dust around
the emitting volume. If we adopt 
the reddening law by Rieke \& Lebofsky (1985), then the {\em variation} 
of both $JHK$ magnitudes and NIR colors depends only on the temperatures  
of the two regions and their opacities (in the $K$ band). 
We checked that, by suddenly heating up the outer layer,
a magnitude decrease of $\sim 1$ mag (in the $JHK$ 
bands) and a blueing in $J-H$, $H-K$ of only few tenths of a mag can be yielded. 
This just using  
slightly different temperatures in the two regions (around 1500 K)  
and an optically thin outer envelope ($\tau_{K} \sim 0.1$). It is noteworth also 
that an optically thin inner layer produces better results.  
This would agree with what found by Foley et al.\ (2007), i. e., that 
there is little mass in the CSM, although there is evidence that 
it is quite dense. Our results   
are roughly consistent with both the $JHK$ magnitude decrease  
and the $H-K$ behavior around the re-brightening epoch, but fail  
to fit the apparent reddening of $J-H$ around peak evident in Fig.~\ref{fig3}
(at $H-K \sim 0.8$, $J-H \sim 1.2$). 
We also checked that two thicker layers ($\tau_{K} \sim 0.5$), with the
outer one a little thicker and pre-existing, would fit better the data,
since now $J-H$ is reddened by few tenths of a magnitude (along with
$H-K$, that however is less reddened). This may suggest that a shock
proceeding in a moderately thick shell of circumstellar matter could fit
marginally better the NIR data, although it does not agree with the
estimated limit to the extinction. 
However, our two-layer model is too rough 
to help us deciding which is the best selection of physical parameters.
Accurate modeling including dust properties (the adopted reddening 
law by Rieke \& Lebofsky is likely not to fit newly-formed grains) 
and the actual distribution 
of the circumstellar matter around the SN is then needed to carefully 
investigate the proposed processes. 

Sakon et al.\ (2007) observed SN2006jc in the NIR and MIR with IRC, on 
board the AKARI satellite, 200 days after discovery. They showed that the 2--5
$\mu$m spectrum is well fitted by thermal emission from amorphous carbon grains at
$\sim 800$ K. Smith et al.\ (2007) also suggest that some 
observational evidence gathered roughly 79 days after discovery, 
points to hot ($\sim 1700$ K) dust formation (carbonaceous grains) in a dense 
circumstellar shell engulfed by the shock. 
These confirm our scenario of dust cooling from $\sim 2000$ K to $\sim 1000$ K.
According to Sakon et al.\ (2007), their derived temperature 200 days after discovery
is consistent with newly formed dust; the total mass in carbonaceous grains
at that date has been estimated to be $\sim 7 \times 10^{-5}$ $M_{\odot}$.  
This is of the same order of the upper limit to the dust mass we have derived in
Sect.~\ref{ccd:text} and may suggest that dust is not distributed with spherical
symmetry. However, by using the same grain properties as in Sakon et al.\ (2007),
we obtain a slightly
lower opacity (in the $K$ band) than our estimated limit even in spherical symmetry. 

\begin{figure}  
\includegraphics[angle=-90,scale=.65]{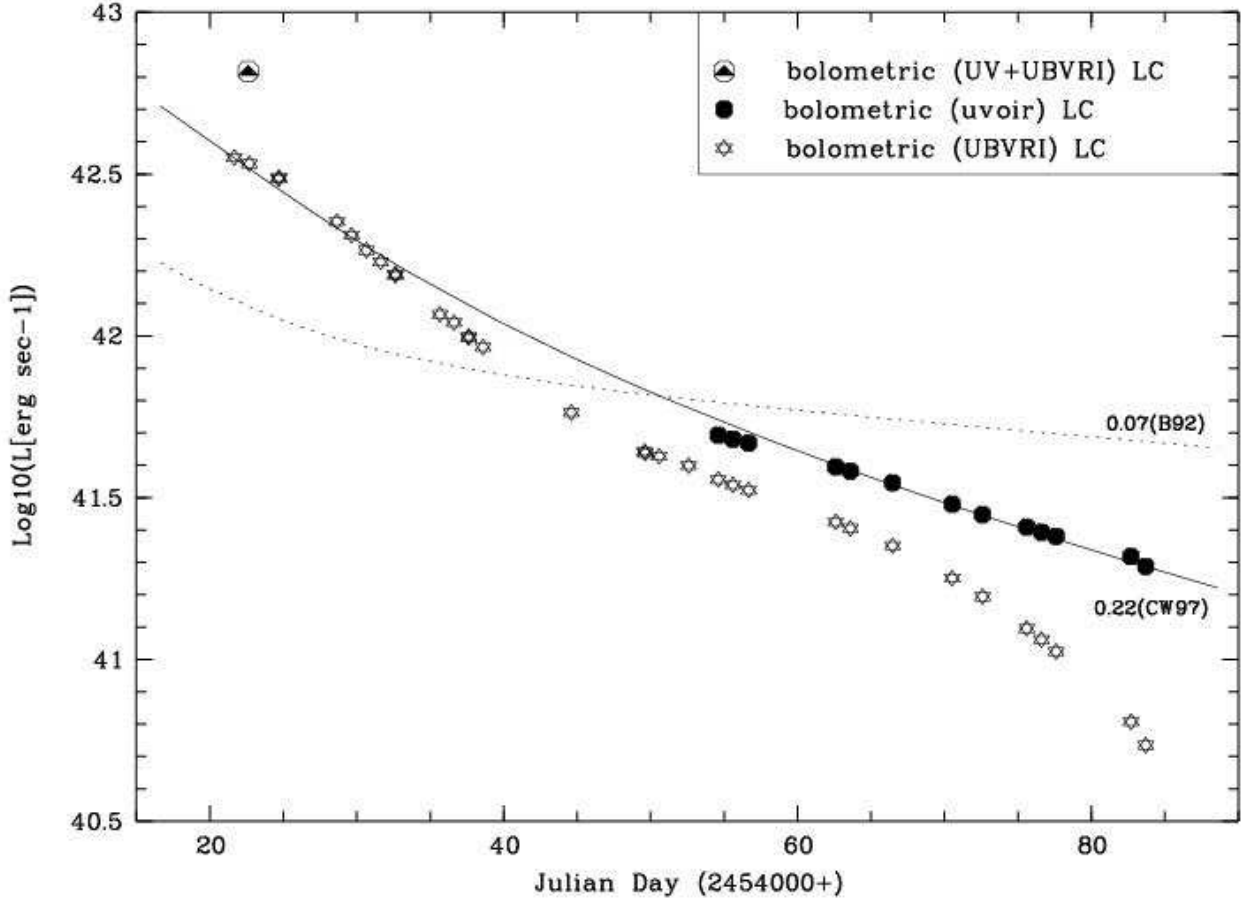}    
\caption{``uvoir'' bolometric light curve and $UBVRI$ (quasi-)bolometric light curve, complementing the NIR
data with optical photometry from Pastorello et al.\ (2007). The topmost
point includes also UV fluxes from Brown et al.\ (2007) to illustrate the
UV contribution to the bolometric luminosity (see the text). Also plotted:
the radioactive decay curve of $^{56}$Ni and $^{56}$Co (dotted line; Branch 1992)
for an initial mass $0.07$ $M_{\odot}$ of $^{56}$Ni; the same curve
(solid line), but
for an initial mass $0.22$ $M_{\odot}$ of $^{56}$Ni and modified according to
Clocchiatti \& Wheeler (1997) to account for varying $\gamma$-ray trapping
(by assuming a kinetic energy $20 \times 10^{51}$ erg s$^{-1}$ and an ejecta
mass $4.6$ $M_{\odot}$). In both case, we adopt JD2454005 as the explosion
time. \label{bolo}}  
\end{figure}

Our NIR data alone do not allow us to discern
between newly formed dust in dense post-shock gas and heating of pre-existing
dust. But this issue can be addressed by comparing optical and NIR observations.
In Fig.~\ref{bolo}, the curve of luminosity from the
radioactive decay of $^{56}$Ni and $^{56}$Co has been drawn (Branch 1992) for
an initial mass $\sim 0.07$ $M_{\odot}$ of $^{56}$Ni. This process is believed to power
the light curve at late epochs. No wonder that radioactive decay does not fit
the slopes of the uvoir and $UBVRI$ (quasi-)bolometric curves, since an increasing
escape probability of gamma rays with time is expected (Clocchiatti \& Wheeler 1997). 
A simple model accounting for this effect in an expanding envelope has been
developed by Clocchiatti \& Wheeler (1997); it can be seen in Fig.~\ref{bolo} that
by modifying the radioactive decay curve by including the decrease of $\gamma$-ray trapping,
we obtain a good fit to the uvoir bolometric curve from JD2454060 on. Here we have assumed
an initial mass $\sim 0.22$ $M_{\odot}$ of $^{56}$Ni, a kinetic energy $\sim 20 \times 10^{51}$
erg s$^{-1}$, an ejecta mass of $4.6$ $M_{\odot}$ and JD2454005 as the explosion date.
These values are the same as the parameters derived by Tominaga et al.\ (2007)
through one of their explosion models.
However, note that our uvoir curve does not include the UV contribution, that appears to
be a consistent fraction of the total radiated energy.

We draw attention to the clear steepening of the $UBVRI$ (quasi-)bolometric curve around
JD2454070, after being roughly parallel to the modeled bolometric curve for $\sim 20$ days. This is 
obviously not matched by the uvoir bolometric curve and reflects an analogous
steepening of the $U,B,V,R,I$ LCs around the same date (see Fig.~\ref{fig3}). 
This has not been found in other 
Ib/c SNe, where the decline rate both of the $V$ band LC and of the $UBVRI$ 
(quasi-)bolometric curve, although larger than the slope of radioactive
decay curve, remains constant until later epoch (Elmhamdi et al.\ 2004, Pastorello et al.\ 2007,
Valenti et al.\ 2007). I. e., they usually are correlated to the radioactive decay
curve (when $\gamma$-ray trapping is accounted for) until late epochs. The 
drop of the $UBVRI$ curve is riminiscent
of what found for SN1990I significantly later, around 250 days after maximum, 
and attributed to dust formation (Elmhamdi et al.\ 2004). 
Since in the case of SN2006jc the $UBVRI$ drop is well correlated with the re-brightening,
in our view it also is the signature of incipient dust formation, occurring earlier than
in the case of SN1990I. 
The drop of the $UBVRI$ (quasi-)bolometric curves is also evident in Fig.~9 of
Tominaga et al.\ (2007), where it clearly departs from their modeled bolometric curves
from the same date on. Note that the analogous drop in the $I$ band LC is less than that in the $V$ band LC
(see Fig.~\ref{fig3}),
as is expected from the extinction law. The amount of extinction at 
maximum $K$ can be roughly derived 
by fitting a linear relation to the $UBVRI$ (quasi-)bolometric curve in the period 
JD2454050--2454070 and measuring the difference between the actual curve and this
line. This implies an $A_{V} \sim 0.5$ mag, or $\sim 3.5 \times 10^{-6}$ $M_{\odot}$
of dust assuming it is homogeneously distributed in a spherical volume of radius
$R$ (as defined in Sect.~\ref{ccd:text}). This values are much lower than those involved in our pure extinction scenario,
then consistent with its rejection.

The last open issue is that of the source of energy fueling the re-brightening.
As shown, the uvoir bolometric curve appears to be consistent with the curve
of radioactive decay, once modified by including the simple $\gamma$-ray trapping
model developed by Clocchiatti \& Wheeler (1997). This would accord with Tominaga et
al.\ (2007), that propose the radioactive decay as the major energy input
powering the LCs. Nevertheless, the presence of UV excess in two spectra taken during the 
re-brightening (Smith et al.\ 2007) would indicate the shock as the major contributor to
grain heating, leading to the NIR re-brightening. This is also suggested by 
the concomitant brightening of X-ray emission,
from $\sim 20$ days to $\sim 100$ days after discovery (Immler et al. 2007).
Holland et al.\ (2007) observed SN2006jc through the three UV bands (UVW1, UVM2, UVW2)
allowed by UVOT on board the Swift satellite. Although still tentative, their data seem to show
a decrease in the slope of the corresponding LCs during the re-brightening and they found
evidence that the UV color became bluer with time. As shown in Sect.~\ref{blc:text},
the UV to the optical {\em plus} NIR luminosity ratio appears to remain constant with time;  
since the NIR increases in fraction with respect to the optical, then the UV fraction has 
also to increase with respect to the optical. Pending a refinement of data
reduction, the available UV and X-ray data point to a correlation with the 
NIR re-brightening that would confirm the occurring of an interaction of the ejecta
with circumstellar matter.

In summary, the case for a strong interaction of SN2006jc ejecta and 
a dense circumstellar medium leading to the formation of hot dust 
in the post-shock denser gas,
appears well supported by the available dataset of multi-wavelength observations. 
\\

\acknowledgments  
We wish to thank Amedeo Tornamb\'{e} for discussions, suggestions and 
encouragement at all stages of this work.
We thank also M.\ Limongi and N.\ Tominaga for their kindly collaboration. \\
This work is supported by governmental grant PRIN-MIUR 2006, under the 
scientific project: {\it ``Fasi finali dell'evoluzione stellare - Nucleosintesi
in Supernovae.``}

This publication makes use of data products from the Two Micron All Sky
Survey, which is a  joint project of the University of Massachusetts and the
Infrared Processing and Analysis Center/California Institute of Technology,
funded by the National Aeronautics and Space Administration and the National
Science Foundation.   \\

\clearpage

\clearpage     

\clearpage

\clearpage

\clearpage  
 
\clearpage

\begin{deluxetable}{ccccccc}  
\tabletypesize{\footnotesize}  
\tablecaption{NIR photometry of SN2006jc. \label{table1}}  
\tablewidth{0pt}  
\tablehead{  
\colhead{Julian Day} & \colhead{$J$} & \colhead{$\Delta J$} & \colhead{$H$} &  
\colhead{$\Delta H$} & \colhead{$K$} & \colhead{$\Delta K$}   \\   
\colhead{(2454000+)}& \colhead{(mag)} & \colhead{(mag)} & \colhead{(mag)} &   
\colhead{(mag)} & \colhead{(mag)} & \colhead{(mag)}  }  
\startdata 
  54.6 &   16.16 & 0.04    &   15.68 &  0.04    &   15.12 &  0.03   \\  
  55.6 & \nodata & \nodata &   15.64 &  0.04    & \nodata & \nodata \\   
  62.6 & \nodata & \nodata & \nodata &  \nodata &   14.74 &  0.04   \\  
  63.6 &   16.03 & 0.05    &  15.54  &  0.03    & \nodata & \nodata \\   
  72.6 &   16.12 & 0.05    &  15.24  &  0.02    &  14.45  &  0.04   \\  
  75.6 &   16.05 & 0.04    &  15.07  &  0.02    &  14.27  &  0.02   \\  
  76.6 &   16.09 & 0.05    & \nodata & \nodata  & \nodata & \nodata \\  
  77.6 & \nodata & \nodata &  15.00  &  0.02    &  14.17  &  0.02   \\  
  82.7 &   15.95 & 0.06    &  14.78  &  0.02    &  13.96  &  0.02   \\  
  83.7 &   15.96 & 0.04    &  14.75  &  0.02    &  14.03  &  0.02   \\  
 132.3 &   17.67 & 0.17    &  15.99  &  0.03    &  14.78  &  0.02   \\  
 137.3 &   17.64 & 0.15    &  16.23  &  0.03    &  14.94  &  0.03   \\  
 159.4 &   18.18 & 0.31    & \nodata &  \nodata &  15.22  &  0.04   \\  
 162.4 & \nodata & \nodata & 16.75   & 0.05     &  15.46  &  0.10   \\  
 164.4 &   18.77 &  0.38   & 16.83   & 0.03     &  15.32  &  0.05   \\  
 165.5 &   18.74 &  0.33   & \nodata & \nodata  &  15.35  &  0.05   \\  
 168.4 & \nodata & \nodata & \nodata & \nodata  &  15.42  &  0.06   \\  
 172.4 &   18.90 & 0.40    &  16.80  &  0.02    &  15.48  &  0.05   \\  
 173.4 & \nodata & \nodata &   16.87 &  0.02    &  15.43  &  0.06   \\  
 174.4 & \nodata & \nodata &   16.89 &  0.02    &  15.30  &  0.07   \\  
 175.4 &   18.84 &  0.48   &  17.0 8 &  0.04    &  15.62  &  0.12   \\  
 176.4 & \nodata & \nodata &   16.96 &  0.03    &  15.54  &  0.11   \\  
 177.4 & \nodata & \nodata &   17.04 &  0.02    &  15.61  &  0.09   \\  
 181.3 & \nodata & \nodata & \nodata & \nodata  &  15.58  &  0.11   \\  
 197.4 &   18.86 & 0.50    &  17.45  &  0.17    &  15.91  &  0.09   \\  
 202.3 &   18.98 & 0.51    &  17.78  &  0.28    &  16.13  &  0.11   \\  
 207.3 &   19.06 & 0.72    & \nodata & \nodata  &  16.25  &  0.13   \\  
 219.3 & \nodata & \nodata & \nodata & \nodata  &  16.51  &  0.11   \\  
 226.3 & \nodata & \nodata & \nodata & \nodata  &  16.53  &  0.11   \\  
 235.2 & \nodata & \nodata & \nodata & \nodata  &  16.68  &  0.11   \\  
 258.3 & \nodata & \nodata & \nodata & \nodata  &  17.58  &  0.21   \\  
\enddata  
\end{deluxetable}   

\clearpage  

\begin{deluxetable}{ccc}  
\tabletypesize{\footnotesize}    
\tablecaption{Bolometric luminosities for SN2006jc. \label{table2}}  
\tablewidth{0pt}  
\tablehead{  
\colhead{Julian Day} & \colhead{$Log_{10}(L_{UBVRI})$} & 
\colhead{$Log_{10}(L_{uvoir})$}  \\   
\colhead{(2454000+)}& \colhead{[erg s-1]} & \colhead{[erg s-1]}}  
\startdata                             
 21.66  &    42.55   & \nodata  \\ 
 22.68  &    42.53   & \nodata  \\ 
 24.68  &    42.48   & \nodata  \\ 
 24.70  &    42.49   & \nodata  \\ 
 28.65  &    42.35   & \nodata  \\ 
 29.64  &    42.31   & \nodata  \\ 
 30.66  &    42.26   & \nodata  \\ 
 31.63  &    42.23   & \nodata  \\ 
 32.63  &    42.19   & \nodata  \\ 
 32.63  &    42.19   & \nodata  \\ 
 35.66  &    42.07   & \nodata  \\ 
 36.62  &    42.04   & \nodata  \\ 
 37.61  &    42.00   & \nodata  \\ 
 37.62  &    41.99   & \nodata  \\ 
 38.59  &    41.96   & \nodata  \\  
 44.60  &    41.76   & \nodata  \\ 
 49.58  &    41.64   & \nodata  \\  
 49.64  &    41.64   & \nodata  \\ 
 50.58  &    41.63   & \nodata  \\  
 52.59  &    41.60   & \nodata  \\  
 54.60  &    41.56   &  41.69   \\ 
 55.60  &    41.54   &  41.68   \\
 56.67  &    41.52   &  41.67   \\
 62.60  &    41.43   &  41.59   \\
 63.60  &    41.41   &  41.58   \\
 66.47  &    41.35   &  41.55   \\
 70.53  &    41.25   &  41.48   \\
 72.60  &    41.19   &  41.45   \\
 75.60  &    41.10   &  41.41   \\
 76.60  &    41.06   &  41.39   \\
 77.60  &    41.02   &  41.38   \\
 82.70  &    40.81   &  41.32   \\
 83.70  &    40.73   &  41.29   \\
\enddata  
\end{deluxetable}   

\end{document}